\documentclass[twocolumn,showpacs,preprintnumbers,prl]{revtex4}

\usepackage[dvips]{graphicx}
\usepackage[dvips]{color}

\usepackage{graphicx}
\usepackage{epsfig}

\begin{document}
\title{Selective Dynamic Nuclear Spin Polarization in Spin-Blocked Double-Dot}
\author{Changxue Deng}
\author{Xuedong Hu}
\affiliation{Department of Physics, University at Buffalo, SUNY,
Buffalo, NY 14260-1500}
\date{\today}

\begin{abstract}
We study the mechanism of dynamical nuclear spin polarization by hyperfine
interaction in spin-blocked double quantum dot system.  We calculate the
hyperfine transition rates and solve the master equations for the nuclear
spins.  Specifically, we incorporate the effects of the nuclear quadrupole
coupling due to the doping-induced local lattice distortion and strain.  Our
results show that nuclear quadrupole coupling induced by the 5\% indium
substitution can be used to explain the recent experimental observation of
missing arsenic NMR signal in the spin-blocked double dots.  
\end{abstract}
\pacs{85.35.Be, 76.60.-k, 03.67.Lx, 
}
\maketitle

In many metals and semiconductors, nuclei and conduction electrons form a
coupled nonlinear spin system through the hyperfine interaction.  Such a
coupling is well studied in bulk experimental phenomena such as dynamical nuclear
spin polarization \cite{OO}.  
Both electron and nuclear spins in semiconductor heterostructures have been
proposed as candidates for qubits in a scalable quantum information
processor.  There are also a variety of spintronic devices where electron
spins provide novel functionalities.  For both quantum computation and
spintronics, understanding the electron-nuclear spin interaction and the
resulting coupled dynamics is crucial because of its important roles in spin
manipulation 
and decoherence.  
%
%
Indeed, in the past decade there have been extensive experimental and
theoretical studies of electron-nuclei hyperfine interaction in semiconductor
nanostructures \cite{Gammon-science, Kawa, decoherence, Zoller, SB-NS}.  In
this Letter, we show that nuclear quadrupole coupling can be a critical
factor in this coupled dynamics, thus needs to be carefully accounted for in
the study of spin-based quantum computers and spintronic devices.


Dynamical nuclear spin polarization has recently been demonstrated in a spin
blocked semiconductor double quantum dot made of Ga$_{0.95}$In$_{0.05}$As
\cite{SB-NS,SB,QD-rev}.  Coulomb interaction and Pauli principle dictates that
in a double quantum dot two-electron singlet and triplet states are split, so
that proper voltage offset and bias between the double dot leads to
significant suppression in the tunnel current due to occupied triplet state
(thus the so-called spin blockade regime, which has been suggested for single
spin detection, a crucial component of quantum information processing)
\cite{SB}.  One way to lift this blockade is to apply an appropriate magnetic
field,
%
%
so that electron-nuclear spin hyperfine coupling can facilitate electron
spin-flip transitions, which in turn leads to dynamical nuclear spin
polarization as was shown in Ref.~\cite{SB-NS}.  The reported experimental
observations reveal several quite unexpected phenomena \cite{SB-NS}, one of
which is that 
%
%
while nuclear magnetic resonance (NMR) signals were observed at the
$^{69}\!$Ga and $^{71}\!$Ga frequencies, no comparable response has been
found from the $^{75}\!$As nuclei.  It seems that nuclear spin polarization
cannot be built up in $^{75}\!$As nuclei even though they have the largest
concentration and the largest hyperfine coupling with the electrons among the
three nuclear species.  

In this Letter we study the mechanism of dynamical nuclear spin polarization
by hyperfine interaction in a spin-blocked double quantum dot, taking into
account electron tunneling and nuclear quadrupole coupling.  We show that the
absence of $^{75}\!$As nuclear spin signal is due to the local strain created
by the presence of Indium atoms.  Specifically, the substitution of the 5\%
In for Ga breaks the crystal symmetry and induces a local lattice distortion
\cite{Quad0,Quad}.  The charge redistribution generates {\it static} electric
field gradients (EFG) at the surrounding lattice sites, which lead to finite
quadrupole coupling for the nuclei \cite{GaAs_no_quad}.  The randomly
distributed In atom only substitute one of the nearest neighbors of
As atoms, while the nearest neighbors of Ga atoms are always four As atoms. 
Therefore the local electric field gradients at the Ga nuclei sites are on
average much smaller than at the As sites.  We show that this difference of
EFG at the locations of As and Ga nuclei is the major origin of the
unexpected absence of As nuclear spin signal.  

We first note that the static quadrupole interaction alone cannot generate any
transition among the spin eigenstates because there is no energy relaxation
mechanism.  This is different from the spin-lattice effect, where the EFG
is generated by the lattice oscillations.  There quadrupole spin relaxation
can be realized by absorbing and emitting a phonon simultaneously (i.e., a
Raman process) \cite{spin-lattice}.  In the system we consider, the nuclei
couple with the electrons, which in turn couple to the leads (reservoir)
by tunneling, so that energy conservation can be satisfied through charge
exchange between the dot and the leads.  
%
%
In the following we first find the nuclear spin eigenstates including the
static quadrupole interaction, then evaluate the first-order transition rates
among these eigenstates due to hyperfine coupling.  
%

The static Zeeman and quadrupole Hamiltonian for a single nuclear spin is
\begin{equation}
H_{n} = -\omega \hat{I}_z + \nu ( \hat{I}_Z^2 - \frac{1}{3} \hat{I^2} ),
\label{eq-H_n}
\end{equation}
where $\omega$ is the nuclear spin Lamour frequency and $\nu$ is quadrupole
coupling strength that is proportional to the EFG at the locations of the
nuclei, $z$ is the direction of external magnetic field and $Z$ is the
principle axis of the largest EFG \cite{NMR}.  We have
neglected the asymmetrical effect of quadrupole interaction.  It was
determined experimentally that $\nu_{\text{As}} \approx \text{2.5 MHz}$ and
$\nu_{\text{Ga}}$ is much less than $\nu_{\text{As}}$ in In$_x$Ga$_{1-x}$As
for $x$ up to 0.01 \cite{Quad}.  For a typical external field $B_{\text{ext}}$ (0.8 Tesla)
applied in the experiment \cite{SB-NS}, we can assume $\nu_{\text{Ga}} \ll
\omega_{\text{Ga}} \approx 10.4 ~\text{and} ~8.2 ~\text{MHz}$ for $^{69}$Ga
and $^{71}$Ga nuclei.  Since our intention is to study the relation of As and
Ga nuclear polarization, we will not differentiate $^{69}$Ga and $^{71}$Ga
further in this Letter.  The small $\nu_{\text{Ga}}$ allows eigenstates $|\phi_i \rangle$
($i$=-3/2,-1/2,1/2 or 3/2) of $H_n$ for Ga nuclei to be expressed
perturbatively in terms of the Zeeman spin eigenstates $|m \rangle$.
\begin{eqnarray}
| \phi_{\pm \frac{3}{2} } \rangle &=& |\pm \frac{3}{2} \rangle - b ~|\pm
\frac{1}{2} \rangle \mp c ~|\mp \frac{1}{2}\rangle , \nonumber \\
| \phi_{\pm \frac{1}{2} } \rangle &=& |\pm \frac{1}{2} \rangle + b ~|\pm
\frac{3}{2} \rangle \mp c ~|\mp \frac{3}{2} \rangle ,
\label{eq:eigen1}
\end{eqnarray}
where $b=\sqrt{3}\nu_{\text{Ga}}\text{sin}\theta \text{cos}\theta
/\omega_{\text{Ga}}$ and $c = \sqrt{3}\nu_{\text{Ga}}\text{sin}^2\theta
/4\omega_{\text{Ga}}$.  $\theta$ is the angle between $z$ and $Z$ axes.  In
these calculations $x$ axis is chosen to be in the plane of $z$ and $Z$. 
These new states $|\phi_i \rangle $ are not eigenvectors of $\hat{I}_z$
anymore, though their average spin polarizations $\langle \phi_i | \hat{I}_z |
\phi_i \rangle = i + {\it O}(\nu_{\text{Ga}}^2/\omega_{\text{Ga}}^2)$.


Transitions can be induced by the hyperfine interaction
among the new energy eigenstates of the Ga nuclei.  The hyperfine
Hamiltonian is
\begin{equation}
H_h = \sum_i^{N} A_i |\psi({\bf R}_i)|^2 \left[ \hat{I}_z^{i} \hat{S}_z +
\frac{1}{2}(\hat{I}_+^{i} \hat{S}_- + \hat{I}_-^{i} \hat{S}_+ ) \right].
\label{eq:hyper}
\end{equation}
Here $N \sim 10^5$ is the number of nuclei in a quantum dot; $A$ is the
hyperfine coupling constant; $\psi({\bf R}_i)$ is the electron envelope
wavefunction at the $i$th nucleus.  Using the electron density at nuclear
sites \cite{Paget} we estimate $A \approx 46~\mu$eV  and 40 $\mu$eV for 
As and Ga nuclei respectively.  Hamiltonian~(\ref{eq:hyper}) describes the interaction of an
electron with an ensemble of $N$ nuclei in the dot.  

To calculate the transition rates given in Fig.~1 we must study how the system
couples to the environment to satisfy energy conservation.  In the
present case electron exchange between a dot and the leads is the most
important energy relaxation mechanism \cite{NSR}.  For triplet state
$|\downarrow \downarrow \rangle$ \cite{note_ST} the electron in the 1st (2nd) dot only
couples to the left (right) lead.  The electron tunnel Hamiltonian can be
expressed as
\begin{equation}
H_e = \sum_{k} \epsilon_k c_{k \downarrow}^+c_{k \downarrow}
+ \epsilon_0 d_{\downarrow}^{+} d_{\downarrow}
+ \sum_{k} \left [t_k c_{k \downarrow}^{+} d_{\downarrow} + \text{H.C.}
\right],
\label{eq-He}
\end{equation}
where $\epsilon_0$ is the renormalized single particle energy level.  
Since the Hamiltonian
$H_e$ is quadratic, exact solution of the retarded Green's function
$G_{\downarrow \downarrow}^{Ret} = -i \theta (t) \langle \{d_{\downarrow}(t),
d_{\downarrow}^{+}(0)\} \rangle$ can be obtained by using the equation of
motion method.  The density of state (DOS) follows from $\rho_T (\epsilon) =
-\text{Im} G_{\downarrow \downarrow}^{Ret} (\epsilon + i0^+)/ \pi$.  It is
straightforward to show that the electron triplet state is broadened
\begin{equation}
\rho_T(\epsilon) =  \frac{1}{2\pi} \frac{\Gamma}{(\epsilon - \epsilon_0)^2 +
\Gamma^2/4},
\label{eq:dos}
\end{equation}
where $\Gamma = 2\pi \sum_k |t_k|^2 \delta(\epsilon-\epsilon_k)$ is the level 
broadening for the triplet state. 
%
%
For the electron in the singlet state, the DOS should be
modified by a weak inter-dot coupling.  However, the major energy relaxation
channel is still the lead-dot tunnel coupling, so we can approximate the
singlet DOS with Eq.~(\ref{eq:dos}) as well. 

\begin{figure}[t]
\begin{center}
\epsfig{file=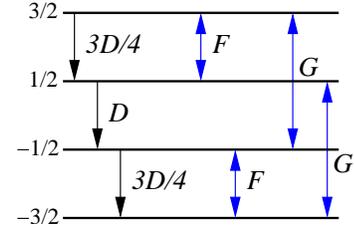, width=4.5cm,height=3cm}
\caption{Schematic of hyperfine induced transitions among the energy
eigenstates of Hamiltonian $H_n$.  The downward arrows represent simultaneous
flip-flop scattering of nuclei and the electron in a triplet state.  $F$ and
$G$ are the nuclear spin relaxation rates induced by the effective hyperfine
magnetic field of the electron ($\frac{A}{N} I_z S_z$).}
\label{fig1}
\end{center}
\end{figure}

The first term in Eq.~(\ref{eq:hyper}) does not involve electron spin flip. 
However, it can induce transitions among the energy eigenstates due to 
mixture of different spin eigenstates in Eq.~(\ref{eq:eigen1}). The
transition rate between state $|\phi_i \rangle$ and $|\phi_j \rangle$
due to this term is given by the Fermi golden rule,
\begin{equation}
W_{i,j}^{z} = \frac{\pi A^2}{2\hbar N^2} |\langle \phi_i | \hat{I}_z | \phi_j
\rangle |^2 P(E_j - E_i),
\label{eq:Wz}
\end{equation}
where $E_i$ and $E_j$ are the eigenenergies of $H_n$.  The absorption power
$P$ is
\begin{equation}
P(\Delta) = \sum_{\epsilon_i,\epsilon_f} \rho (\epsilon_i)
\rho (\epsilon_f) \delta(\epsilon_f - \epsilon_i + \Delta),
\label{eq:P}
\end{equation}
where $\rho(\epsilon)$ is the DOS of either the electron singlet or the
triplet state.  Using Eq.~(\ref{eq:Wz}), the transition rates in
Fig.~(\ref{fig1}) can be calculated: $F$ = $6b^2A^2/(N^2 \hbar \Gamma)$ and
$G$ = $6c^2A^2/(N^2 \hbar \Gamma)$.  In deriving these expressions we have
neglected the nuclear energy difference of the initial and final state since
they are much smaller than the width of electronic energy broadening.  

The calculation of the transition rates due to the hyperfine flip-flop terms
in Eq.~(\ref{eq:hyper}), which correspond to the polarized triplet to singlet
electron transitions, is more complicated.  When the two electron states are
degenerate, the hyperfine flip-flop transition is on resonance, leading
to nuclear spin polarization.  However, the transition should quickly becomes off
resonance due to the effective nuclear magnetic field $B_{\text{N}}$
experienced by the electrons. 
%
%
The effective nuclear magnetic fields can be estimated as
\begin{equation}
B_{\text{N}} = \gamma_{\text{As}} \frac{\langle \hat{I}_z \rangle
_{\text{As}}}{I} + \gamma_{\text{Ga}} \frac{\langle \hat{I}_z \rangle
_{\text{Ga}}}{I},
\label{eq:BN}
\end{equation}
where $\gamma_{\rm As} \approx -2.8 $ Tesla and $\gamma_{\rm Ga} \approx -1.3$
Tesla \cite{Paget}.  As and Ga nuclei have independent spin temperatures
because their mutual flip-flop process is largely suppressed due to different
magnetic moments.  Equilibrium in nuclear spin polarization will thus be
established in each of the nuclear species independently.  $B_{\rm N}$ can be
quite large for even moderate nuclear polarization.  The energy difference of
the singlet state and triplet state due to $B_{\text{N}}$ can be approximated
by $\Delta_{ST} = -g^*\mu_B B_{\text{N}}$ assuming the energy of the singlet
state is independent of magnetic field.  The flip-flop transition rates of
hyperfine interaction from state $|\phi_i \rangle $ to state $|\phi_j
\rangle$ is thus
\begin{equation}
W_{i,j}^{-} = \frac{f_T \pi A^2}{8\hbar N^2}
| \langle \phi_{j} | \hat{I}_{-} | \phi_i \rangle |^2 P(\Delta_{ST} + E_{j} -
E_i),
\label{eq:W-}
\end{equation}
where $f_T$ is a multiplication factor due to cotunneling and the
probability of the triplet state occupation (electrons can tunnel into three 
triplet states and one singlet state, but only one triplet is relevant to 
nuclear polarization). $f_T$ should be multiplied by another factor 0.5 because 
there are two dots and the probability of nuclear spin-flip process in each of the dots is
50\%. We estimate $f_T \approx 0.05$.  
%
%
We then find that $D$ in Fig.~(\ref{fig1}) is given as $D = f_T A^2 \Gamma
/N^2\hbar (\Delta_{ST}^2 + \Gamma^2)$.

The quadrupole interaction of As nuclei is much stronger than that of Ga
[$\nu_{\text{As}} \approx 2.5$ MHz is in the same order as the Zeeman energy
$\omega_{\text{As}} \approx 5.9$ MHz (at 0.8 Tesla)] and leads to a complete
mixing of the spin eigenstates, so that the perturbative wavefunctions
[Eq.~(\ref{eq:eigen1})] for Ga nuclei have to be replaced with numerical
solutions:
%
%
\begin{equation}
|\phi_i \rangle = \sum_{m} u_{im} |m \rangle.
\label{eq:eigen2}
\end{equation}
The expectation value of $\hat{I}_z$ of these state is $\langle \phi_i |
\hat{I}_z | \phi_i \rangle = \sum_{m} |u_{im}|^2 m$.  For example, we find
that when $\alpha \equiv \nu_{\text{As}} / \omega_{\text{As}} =0.3$, 
$\langle \phi_i | \hat{I}_z | \phi_i \rangle $ = 1.419, 0.558, -0.528 and
-1.450 for $i$ = 3/2, 1/2, -1/2 and -3/2 respectively.  All the matrix
elements of $\hat{I}_z$ and $\hat{I}_-$ can be easily obtained from
Eq.~(\ref{eq:eigen2}), and the transition rates can be calculated numerically
using Eq.~(\ref{eq:Wz}) and (\ref{eq:W-}).
%
%
One should keep in mind that the illustration of the various transitions in
Fig.~(\ref{fig1}) is only applicable to the perturbative analysis for Ga
nuclei, while for As nuclei all transitions are possible.  However the
concept of polarization ($D$) and depolarization ($F$ and $G$) processes
illustrated in Fig. (\ref{fig1}) should still be valid.

After obtaining all the transition rates $W^z$ and $W^-$ we can proceed to
construct the master equations \cite{NMR}
\begin{equation}
\frac{d p_i} {dt} = \sum_{j \ne i} W_{j,i} p_j - \sum_{j \ne i} W_{i,j} p_i,
\label{eq:rate}
\end{equation}
where $W_{i,j} = W_{i,j}^z +W_{i,j}^-$, and $p_{i}$ is the probability of 
nuclear spins in state $| \phi_{i} \rangle$.  In Eq.~(\ref{eq:rate}) we have
neglected the spin-lattice relaxation, which is very weak at liquid Helium
temperature.  Here we can use the spin temperature approximation
because the nuclear spin build-up time ($\approx$ 100 s) \cite{SB-NS} is much
longer than the transverse spin relaxation $T_2~ (\approx 10^{-4} ~{\rm s})$.
We solve for the steady-state polarization: $d p_i/dt = 0$.  
These equations are highly nonlinear, as the
equilibrium polarizations depend on the transition rates, while to calculate
the transition rates $W^-$ one must know the effective nuclear magnetic
field, which itself depends on the nuclear spin polarization.  We solve the
set of equations self-consistently.  The average nuclear polarization is $P =
\sum_i p_i \langle \phi_i | \hat{I}_z | \phi_i \rangle = \langle \hat{I}_z
\rangle /I$.  We can then calculate the effective nuclear magnetic fields
$B_{\text{N}}$ with Eq.~(\ref{eq:BN}).  

\begin{figure}[t]
\begin{center}
\epsfig{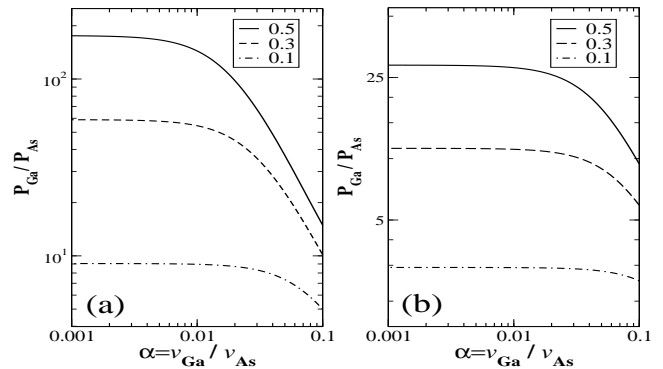}
\caption{ Ratio of Ga and As nuclei polarization as a function of the relative
quadrupole interaction strength $\alpha ~(=\nu_{\text{Ga}}/\nu_{\text{As}})$
for $\Gamma=10~\mu$eV (panel a) and $\Gamma=30~\mu$eV (panel b).
$\Gamma$ is the width of level broadening due to electron tunneling between
the lead and the dot. The three curves in each graph represent three different
As quadrupole coupling strengths $\nu_{\text{As}}/\omega_{\text{As}} $ 
= 0.1, 0.3 and 0.5. $\theta$ has chosen to be 45 degrees for both As and Ga
nuclei.
}
\label{fig2}
\end{center}
\end{figure}

Figure~\ref{fig2} shows the calculated ratio of Ga and As nuclei polarization
as a function of $\alpha$, which is taken as a free parameter in our
calculations.  We consider As quadrupole effects with three different
strengths.  $\nu_{\text{As}}/\omega_{\text{As}} = 0.3$ corresponds to
$\nu_{\text{As}} \approx 1.8$ MHz.  In this case the Ga nuclear polarization
is two orders of magnitude greater than that of As nuclei when $\alpha
\approx$ 0.01.  This difference can qualitatively explain the experimental
observation that the As NMR signal is missing in measuring the current
oscillation period as a function of the AC magnetic field frequency.  We also
find the ratio is very sensitive to electron level width $\Gamma$. 
Figure~\ref{fig2} shows that the polarization ratio in the case of
$\Gamma=30~\mu$eV is roughly $\frac{1}{5}$ of that in the case of
$\Gamma=10~\mu$eV.  Since $D \propto \Gamma/(\Delta_{ST}^2 + \Gamma^2 )$, the
triplet to singlet transition rate decreases much faster when $\Gamma$ is
smaller, resulting in larger polarization ratio.  

\begin{figure}[t]
\begin{center}
\epsfig{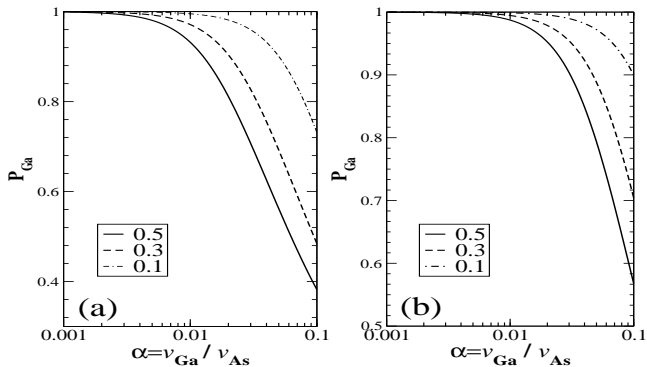}
\caption{ Ga nuclear spin polarization as a function of the relative
quadrupole interaction strength $\alpha$ for $\Gamma=10~\mu$eV (panel a) and
$\Gamma=30~\mu$eV (panel b). All the parameters are the same as those used in
Fig. 2.}
\label{fig3}
\end{center}
\end{figure}

Figure~\ref{fig3} shows how Ga nuclear spin polarization changes with the
relative quadrupole coupling strength $\alpha$.  It is obvious from the two
panels that $P_{\text{Ga}}$ increases rapidly as $\alpha$ decreases from 0.1
to 0.01.  The Ga polarization saturates to unity for smaller $\alpha$, and
larger $\Gamma$ leads to larger Ga polarization.

%
%

The physical picture of the master equation calculation is actually quite
straightforward.  Both Ga and As have a polarization rate and a
depolarization rate.  The As nuclear spin depolarization rate is much larger
than that of Ga due to the quadrupole coupling.  Thus there exists a regime
of polarization rate where nuclear spin polarization can be built up in Ga
but not in As.

%
In our study As depolarization rates are greatly enhanced because the energy
levels of the dots are broadened through tunnel coupling to the leads, and
that the singlet-triplet transition becomes off resonance once the
polarization of Ga nuclei is established.  In previous experiments concerning
dynamical nuclear polarization, optical pumping creates a constant
polarization rate for the nuclei with the help of hyperfine interaction; and
the relaxation process due to carrier recombination overshadows the effect of
quadrupole interaction, so that As polarization can still be built up
\cite{OO}.
 
In conclusion, we have studied the dynamical nuclear polarization mechanism in
spin-blockade double-dot system.  Specifically, we calculate the hyperfine
induced polarization rates and depolarization rates among the eigenstates of
the nuclear spins.  We show that the average spin polarization in steady
state of different nuclear species (Ga and As) could differ by two orders of
magnitude due to static quadrupole interaction induced by lattice
distortions.  Our results can thus be used to explain the recent NMR
experiments conducted in such system.  Our calculation suggests both caution
and promise.  Doping is widely used in semiconductors to tailor electronic
properties.  However, as we point out in this Letter, there can be unexpected
side effects in material properties, such as increased nuclear spin
relaxation in the present case.  Conversely, controlled doping can also be
used to differentiate parts of a system (such as Ga and As nuclei here), so
that selective operations become possible.  How to utilize the additional
control provided by doping while overcoming its negative effects can be
critical to future quantum information processors and/or spintronic devices.

We acknowledge helpful conversations with K. Ono, S. Tarucha, and S. Das
Sarma.  We thank partial financial support by ARDA and ARO.


\begin{thebibliography}{99}
\bibitem{note_ST} We have assumed that the electron g-factor is negative and
the triplet is the ground state at zero magnetic field. One could equivalently
discuss the other triplet state $|1,1 \rangle = |\uparrow \rangle |\uparrow
\rangle $ if the singlet is the ground state when there is no magnetic field
\cite{Jpn}.
\end{thebibliography}
\end{document}